\documentclass[leqno]{article}
\usepackage{jjss-2e,latexsym,graphicx}

\makeatletter

\@addtoreset{equation}{section}
\makeatother
     \def\bm#1{\mbox{\boldmath$#1$}}   \def\cl#1{{\cal #1}}   
\newtheorem{theorem}{THEOREM} \newtheorem{remark}{REMARK}

\volumenumber{99}
\vnumber{99}
\publishyear{9999}
\frompage{9}
\topage{999}

\shorttitle{VARIANCE\n ESTIMATORS\n FOR\n $U$-STATISTICS} 
\jjssvol{Vol.28\n No.1\n 1998}
\title{
DISTURBANCE OF OPERATION IN QUANTUM ESTIMATION FOR THE GAUSSIAN $P$-FUNCTION
}

\author{
Yoshiyuki Tsuda%
\footnote{Imai Quantum Computation and Information Project, ERATO, JST,
5-28-3, Hongo, Bunkyo-ku, Tokyo 113-0033 Japan}\llap,
Keiji Matsumoto${}^*$
and
Masahito Hayashi%
\footnote{Laboratory of Mathematical Neuroscience, Brain Science Institute,
RIKEN, 2-1 Hirosawa, Wako, Saitama 351-0198 Japan}
}%
\abstract{
  For the quantum Gaussian state family, Hayashi proposed
    a quantum mechanical operation using beam splitters to estimate the location
    and scale parameters of the $P$-function, and he showed
    that it is asymptotically optimal.
  In this paper, we analyze the effect of disturbance of his
    operation caused by the randomness of the transparency of the beam splitters.
  It is shown that even if the variance of the random transparency is small,
    Hayashi's estimators are improper in a sense that they are biased and asymptotically inconsistent.
  In such a case, we propose to stop the operation and correct the biases of estimators.
}

\keywords{coherent state, counting measurement, quantum Gaussian state,
heterodyne measurement.}

\begin{document}
\footnotetext{Received March 2003.\m 
Revised Xxx ??, 200X.\m Accepted Yyy, 200X.}

\pagestyle{empty}

\maketitle
\section{Introduction}

The quantum estimation is an application of the statistical estimation theory to the quantum mechanics.
The general theory on the quantum mechanics is described by the linear algebra on a Hilbert space, and it might require some specific setup in order to consider the quantum estimation.
However, our present problem on the Gaussian $P$-function is a special case in which only the basic analytical methods are used.
This paper is unfolding our problem without linear algebraic premises.

\smallskip

The quantum estimation theory  was first considered by Helstrom \cite{helstrom}.
The main problem is to find a optimal scheme to estimate unknown parameter of the quantum state.
Here, the scheme of estimation is composed of (i) physical operation and measurement in order to obtain data and (ii) the estimator based on the obtained data.
Let $\rho$ denote a state of the system to be observed.
Let $M$ denote a measurement on the system.
If one carries out a measurement $M$, then he will obtain data $x$ as an observation of a random variable $X$ whose probability distribution is determined by $\rho$ and $M$.
Suppose that we know the true state is an element of the set $\cl S:=\{\rho_\theta\}$ parameterized by $\theta\in\Theta$, and also suppose that we can select a measurement in the set $\cl M=\{M_\lambda\}$ parameterized by $\lambda\in\Lambda$.
Based on the obtained data $x$, one will estimate $\theta\in\Theta$ by an estimator $\hat\theta$.
In the standard statistical estimation problem, a statistician is allowed to select only the estimator $\hat\theta$.
One features of the quantum estimation is that one could improve the accuracy of the estimation by interaction between samples.
In the present paper, we consider the effect of disturbance of physical operation in a problem of quantum estimation for the Gaussian $P$-function model.

The models of Gaussian $P$-function are typical and important examples of the quantum estimation and it can be applied to the optical experiments.
Our model has two-dimensional location parameters and one scale parameter.
For the estimation problem of the location parameters of the Gaussian $P$-function, Yuen and Lax \cite{yuenlax} and Holevo \cite{holevo} have given a lower bound of the variance of unbiased estimators and an optimal estimation scheme that attains the bound.
For the problem of the location and scale parameters, Hayashi \cite{hayashi} has given a lower bound of the variance of unbiased estimators and an optimal estimation scheme that asymptotically attains the bound.
His result gave the first practical example of the advantage of interaction of samples to the quantum estimation of the Gaussian model concerning the statistical estimation theory.
We consider the model of the location and scale parameters and analyze the effect of disturbance of the asymptotically optimal scheme given by Hayashi \cite{hayashi}.
He pointed out that his scheme is realized in the optical experiment by the physical operation using beam splitters.
For example, if we have given $n=2^m$ samples,
then an experimenter uses beam splitters whose transparency is $1/2$.
However, in practical, the transparency of prepared beam splitters is not always exactly the ideal quantity $(1/2)$, and it is often slightly different from the ideal one.
If the transparency is not correct, then the physical operation proposed by Hayashi \cite{hayashi} is disturbed and it might affect the estimation.
We analyze the effect of the disturbance to the estimation, and it turns out that the estimator is not even asymptotically consistent to the true parameter.
In such a case, a naive scheme not using interactions between samples is better than that of Hayashi \cite{hayashi} since the disturbance could be avoided.
In order to avoid the inconsistency of Hayashi \cite{hayashi}'s estimator, we propose a new scheme to estimate location and scale parameters by correcting that of Hayashi \cite{hayashi}.
We will see that our scheme gives an asymptotically consistent estimator, and it is better than the naive estimator.

\section{Problem}

When we discuss a general problem of quantum estimation, we need to consider any measurement $M$ which may be carried out to the quantum system.
However, our interest in this paper is devoted to the problem of disturbance in the experiment proposed by Hayashi \cite{hayashi}.
His optimal method uses only two kinds of measurements called the 'heterodyne measurement' and the 'counting measurement' and the state is denoted by a probabilistic superposition of the coherent states.
It enables us to describe the problem by a Bayesian model in which the prior distribution of parameters of basic states means the state, and selecting some of random variables corresponds to the selecting measurements.
This is why the linear algebraic preparation is not needed.

Let $(A_j,B_j)$ denote the complex amplitude $A_j+i B_j$ $(i=\sqrt{-1})$ of a coherent state, $(X_j,Y_j)$ denote the observed amplitude $X_j+i Y_j$ by the heterodyne measurement and $Z_j$ denote the counted numbers by the counting measurement.
Suppose that $W_j:=(A_j,B_j,X_j,Y_j,Z_j)$ $(j=1,...,n)$
  are random variables.
Suppose that we can always not observe $A_j$ or $B_j$ for any $j$,
  and we can selectively observe $(X_j,Y_j)$ or $Z_j$ for each $j$,
  so we do not observe $Z_j$ if we select to observe $(X_j,Y_j)$
  and we do not observe $(X_j,Y_j)$ if we select $Z_j$.
Let $S\subseteq\{1,...,n\}$ be a set of indices of which
  we observe $(X_j,Y_j)$, hence we observe $Z_k$ if $k\not\in S$.
Let $(x_j,y_j)$ $(j\in S)$ and $z_k$ $(k\not\in S)$ be observations
  of $(X_j,Y_j)$ and $Z_j$.
For each $j\in S$, we assume that $X_j$ and $Y_j$ are continuous random variables
  on the set $\bm R$ of real numbers,
  and, for each $k\not\in S$ $Z_k$ is a discrete random variable
  on the set $\bm N_0$ of non-negative integers.
We also assume that the joint probability (density) function of
  $(X_j,Y_j)$ $(j\in S)$ and $Z_k$ $(k\not\in S)$ is denoted as
  \[
    E\left[
      \prod_{j\in S}
	\left(
	\frac
	{
		\exp(-(x_j-A_j)^2-(y_j-B_j)^2)
	}
	{\pi}
	\right)
	\prod_{k\not\in S}
	\left(
	\frac
	{
		\exp(-A_k^2-B_k^2) (A_k^2+B_k^2)^{z_k}
	}
	{z_k!}
	\right)
    \right]
  \]
  where the expectation is taken for $A_j$ and $B_j$ $(j=1,...,n)$.

\begin{remark}\label{rm:3:1}\rm
  Suppose that $A_1,...,A_n,B_1,...,B_n$ are independently
    distributed.
  For $j\in S$, $A_j$ and $B_j$ are
    distributed according to the normal distributions
    $N(\theta,\nu/2)$ and $N(\eta,\nu/2)$
    of means $\theta$ and $\eta$ and variance $\nu/2$, respectively.
  For $k\not\in S$, $A_k$ and $B_k$ are
    distributed according to $N(0,\nu/2)$.
  Then, all selected variables $(X_j,Y_j)$ $(j\in S)$ and
    $Z_k$ $(d\not\in S)$ are independently distributed.
  The distributions of $X_j$ and $Y_j$ are $N(\theta,(\nu+1)/2)$
    and $N(\eta,(\nu+1)/2)$, respectively, and
    that of $Z_k$ is the geometric distribution ${\it Geo}(\nu)$
    of mean $\nu$, that has the probability function
    \[
      \frac1{\nu+1}\left(\frac\nu{\nu+1}\right)^{z_k}
      .
    \]
\end{remark}

Not only we can select observed variables, but we can
  also select pairs of variables $(A_j,B_j)$ and $(A_k,B_k)$
  and transform them as
  \[
    \cases
    {
      A_j \ \mapsto \  A_j \cos\tau + A_k \sin\tau, &\cr
      B_j \ \mapsto \  B_j \cos\tau + B_k \sin\tau, &\cr
      A_k \ \mapsto \ -A_k \sin\tau + A_k \cos\tau, &\cr
      B_k \ \mapsto \ -B_k \sin\tau + B_k \cos\tau, &
    }
  \]
  where $\tau\in\bm R$ is arbitrarily selected.
We call this transformation $g_\tau^{(j,k)}$.
Note that the transformations has the group structure in
  the algebraic sense, and we write
  $g_{\tau_2}^{(j_2,k_2)} g_{\tau_1}^{(j_1,k_1)}$ to denote
  that we first carry out $g_{\tau_1}^{(j_1,k_1)}$
  and then next do $g_{\tau_2}^{(j_2,k_2)}$.

\begin{remark}\rm
  Suppose that $A_j,B_j,A_k,B_k$ are independently
    and identically distributed.
  Let $A_j$ and $A_k$ are distributed according to $N(\theta,\nu/2)$
    and $B_j$ and $B_k$ are distributed according to $N(\eta,\nu/2)$.
  If we carry out the transformation $g_\tau^{(j,k)}$,
    then the distributions of $A_j,B_j,A_k,B_k$ are given by
    \begin{eqnarray*}
      &&
      N(\theta(\cos\tau+\sin\tau),\nu/2),\quad
      N(\eta(\cos\tau+\sin\tau),\nu/2),
      \\
      &&
      N(\theta(-\sin\tau+\cos\tau),\nu/2),\quad
      N(\eta(-\sin\tau+\cos\tau),\nu/2),
    \end{eqnarray*}
    respectively.
\end{remark}

Suppose that $A^1,...,A^n,B_1,...,B_n$ are independently distributed,
  and $A_j$ $(j=1,...,n)$ is distributed according to
  $N(\theta,\nu/2)$ and $B_j$ $(j=1,...,n)$ is distributed according to
  $N(\eta,\nu/2)$, where $\theta$, $\eta\in\bm R$ and $\nu>0$
  are unknown.
Under the rules of selecting observed variables and
  transformations, we consider to estimate $\theta$, $\eta$
  and $\nu$ based on observed variables.

\section{Naive estimation without transformation}
In this section, we consider one of the most naive method
  to estimate unknown parameters, that is, we do not
  carry out any transformations and we let $S:=\{1,...,n\}$.
Then, from Remark \ref{rm:3:1}, 
  $X_1,...,X_n,Y_1,...,Y_n$ are independently distributed,
  and $X_j$ is distributed according to $N(\theta,(\nu+1)/2)$
  and $Y_j$ is distributed according to $N(\eta,(\nu+1)/2)$
  for $j=1,...,n$.
Let
  \begin{eqnarray*}
    \bar\theta & := & \frac1n \sum_{j=1}^n X_j,\quad
    \bar\eta   \ := \ \frac1n \sum_{j=1}^n Y_j,
    \\
    \bar\nu & := & \frac1{n-1} \sum_{j=1}^n (X_j-\bar\theta)^2 + \frac1{n-1} \sum_{j=1}^n (Y_j-\bar\eta)^2 -1
  \end{eqnarray*}
  be estimators for $\theta$, $\eta$ and $\nu$.
Then, we have
  \[
    E(\bar\theta) \ = \ \theta,\quad
    E(\bar\eta) \ = \ \eta,\quad
    E(\bar\nu) \ = \ \nu,
  \]
  so that they are unbiased.
Moreover, since the covariance matrix of $(\bar\theta,\bar\eta,\bar\nu)$ is
  \[
    V_1 \ := \
    \pmatrix
    {
      \frac{\nu+1}{2n} & 0 & 0\cr
      0 & \frac{\nu+1}{2n} & 0\cr
      0 & 0 & \frac{(\nu+1)^2}{n-1}
    },
  \]
  we can see that they are asymptotically consistent.

\section{Hayashi's estimation using transformation}
In this section, we consider Hayashi \cite{hayashi}'s method
  to estimate $\theta$, $\eta$ and $\nu$.
Note first that, by some transformations of $A_1,...,A_n,B_1,...,B_n$
  by $G:=g_{\tau_t}^{(j_t,k_t)} \cdots g_{\tau_1}^{(j_1,k_1)}$\llap,
  we obtain random variables $A'_1,...,A'_n,B'_1,...,B'_n$.
  which are mutually independent and
  $A_1,B_1,A_j,B_j$ $(j\ge2)$ are distributed according to
  \begin{equation}
    \label{eq:3:a'}
    N(\sqrt n\theta,\nu/2),\quad N(\sqrt n\eta,\nu/2),\quad
    N(0,\nu/2),\quad N(0,\nu/2),
    ,
  \end{equation}
  respectively.

\begin{remark}\rm
  For $t=1,...,n-1$, let
    \[
      \tau_t \ := \ \tan^{-1} t^{-1/2}.
    \]
  Then, a transformation defined by
    \[
      G_1 \ := \ g_{\tau_{n-1}}^{(1,n)} \cdots g_{\tau_1}^{(1,2)}
    \]
    generates $A'_1,...,A'_n,B'_1,...,B'_n$ which are
    independently distributed according to (\ref{eq:3:a'}).
\end{remark}

If we draw the transformation $g^{(j,k)}_\tau$ as
  \begin{center}
  \begin{picture}(300,40)
    \put(-5,20){$(A_j,B_j)$}\put(40,26){\line(1,0){50}} \put(100,20){$(A_j\cos\tau+A_k\sin\tau,B_j\cos\tau+B_k\sin\tau)$}
    \put(-5, 0){$(A_k,B_k)$}\put(40, 6){\line(1,0){50}} \put(100, 0){$(-A_j\sin\tau+A_k\cos\tau,-B_j\sin\tau+B_k\cos\tau)$}
    \put(65,6){\line(0,1){20}\circle{6}}\put(65,26){\circle*{7}}\put(70,10){$\tau$}
  \end{picture}
  \end{center}
  then, $G_1$ can be expressed as
  \begin{equation}
  \label{fig:3:g1}
  \begin{picture}(180,80)
    \put(0,66){$(A_1,B_1)$}\put(47,68){\line(1,0){100}}\put(155,66){$(A'_1,B'_1)$}
    \put(0,46){$(A_2,B_2)$}\put(47,48){\line(1,0){100}}\put(155,46){$(A'_2,B'_2)$}
    \put(0,26){$(A_3,B_3)$}\put(47,28){\line(1,0){100}}\put(155,26){$(A'_3,B'_3)$}
    \put(0, 6){$(A_4,B_4)$}\put(47, 8){\line(1,0){100}}\put(155, 6){$(A'_4,B'_4)$}
    \put(60,48){\line(0,1){20}\circle{6}}\put(60,68){\circle*{7}}\put(65,53){{$\pi/4$}}
    \put(75,28){\line(0,1){40}\circle{6}}\put(75,68){\circle*{7}}\put(80,33){{$\tan^{-1}2^{-1/2}$}}
    \put(90, 8){\line(0,1){60}\circle{6}}\put(90,68){\circle*{7}}\put(95,13){{$\tan^{-1}3^{-1/2}$}}
  \end{picture}
  \end{equation}
  for $n=4$.

The transformation generating $A'_1,...,A'_n,B'_1,...,B'_n$ which are
  which are distributed according to (\ref{eq:3:a'}) is not
  unique as the follows.

\begin{remark}\label{rm:3:g2}\rm
  Suppose that $n=2^m$ for a natural number $m$.
  For $t=1,...,m$, let $J_t:=\{(j,k)\}$ be the set of pairs of
    natural numbers $j$ and $k$ satisfying
    \[
      j\equiv1\pmod{2^t},\quad
      k-j \ = \ 2^{t-1}
      .
    \]
  Then, a transformation defined by
    \[
      G_2 \ := \ \prod_{(j,k)\in J_m} g_{\pi/4}^{(j,k)} \cdots \prod_{(j,k)\in J_t} g_{\pi/4}^{(j,k)} \cdots \prod_{(j,k)\in J_1} g_{\pi/4}^{(j,k)}
    \]
    also generates $A'_1,...,A'_n,B'_1,...,B'_n$ which are
    independently distributed according to (\ref{eq:3:a'}).
\end{remark}

In the same way as (\ref{fig:3:g1}), $G_2$ can be expressed as
  \begin{center}
  \begin{picture}(180,80)
    \put(0,66){$(A_1,B_1)$}\put(47,68){\line(1,0){100}}\put(155,66){$(A'_1,B'_1)$}
    \put(0,46){$(A_2,B_2)$}\put(47,48){\line(1,0){100}}\put(155,46){$(A'_2,B'_2)$}
    \put(0,26){$(A_3,B_3)$}\put(47,28){\line(1,0){100}}\put(155,26){$(A'_3,B'_3)$}
    \put(0, 6){$(A_4,B_4)$}\put(47, 8){\line(1,0){100}}\put(155, 6){$(A'_4,B'_4)$}
    \put(80,48){\line(0,1){20}\circle{6}}\put(80,68){\circle*{7}}\put(85,53){{$\pi/4$}}
    \put(80, 8){\line(0,1){20}\circle{6}}\put(80,28){\circle*{7}}\put(85,13){{$\pi/4$}}
    \put(113,28){\line(0,1){40}\circle{6}}\put(113,68){\circle*{7}}\put(118,33){{$\pi/4$}}
  \end{picture}
  \end{center}
  for $n=2^2=4$.

After such transformations obtaining $A'_1,...,A'_n,B'_1,...,B'_n$,
  we let $S:=\{1\}$ to select observed variables $(X_1,Y_1),Z_2,...,Z_n$.
Let
  \[
    \hat\theta \ := \ X_1/\sqrt n,\quad
    \hat\eta \ := \ Y_1/\sqrt n,\quad
    \hat\nu \ := \ \frac1{n-1}\sum_{j=2}^n Z_j,
  \]
  then from Remark \ref{rm:3:1}, we have
  \[
    E(\hat\theta) \ = \ \theta,\quad
    E(\hat\eta) \ = \ \eta,\quad
    E(\hat\nu) \ = \ \nu
    ,
  \]
  so that they are unbiased.
Moreover, since the covariance matrix of $(\hat\theta,\hat\eta,\hat\nu)$ is
  \[
    V_2 \ := \
    \pmatrix
    {
      \frac{\nu+1}{2n} & 0 & 0\cr
      0 & \frac{\nu+1}{2n} & 0\cr
      0 & 0 & \nu \frac{\nu+1}{n-1}
    },
  \]
  we can see that they are asymptotically consistent and
  this Hayashi's estimators $\hat\theta,\hat\eta,\hat\nu$
  dominates the naive estimators $\bar\theta,\bar\eta,\bar\nu$
  in a sense that the difference $V_1-V_2$ of the covariance matrices
  is non-negative definite.

Hayashi \cite{hayashi} also proved that his estimators are
  asymptotically optimal in the quantum mechanical setup
  which is more general than that of here.

\section{Noisy transformation}
The transformations $G$ of $A_1,...,A_n,B_1,...,B_n$ to
  $A'_1,...,A'_n,B'_1,...,B'_n$ is closely related to
  physical operation of in quantum optics.
Actually, each $g_\tau^{(j,k)}$ corresponds to the
  interference of two light beams using a beam splitter
  of transparency $\cos^2\tau$.
For example, if an experimenter tries to realize the
  transformation $G_2$, then he has to prepare beam splitters
  of transparency $1/2$, since $G_2$ is constructed
  by $g_{\pi/4}^{(j,k)}$'s.

However, in practical cases, it is difficult to prepare
  beam splitters of exactly the same transparency as the
  ideal quantity $(1/2)$, and in many cases, each quantity
  is slightly different.
Hence, we consider, when $n=2^m$ and we try to carry out
  transformation $G_2$, but each $\tau$ is independently
  and identically distributed according to $N(\pi/4,\epsilon\log2)$,
  and we calculate expectations and variances of Hayashi's estimators
  $\hat\theta,\hat\eta,\hat\nu$ in order to see the effect of
  the randomness of transformations.

Let $\bm A':=(A'_1,...,A'_n,B'_1,...,B'_n)$.
Then, note that Remark \ref{rm:3:1} shows that
  \begin{eqnarray*}
    E(\hat\theta)
    &=&
    \frac1{\sqrt n} E( E( X_1 \mid \bm A ) )
    \ = \
    E(A'_1),
    \\
    E(\hat\theta_2)
    &=&
    \frac1{\sqrt n} E( E( X_2 \mid \bm A ) )
    \ = \
    E(B'_1),
    \\
    E(\hat\nu)
    &=&
    \frac1{n-1} \sum_{j=2}^n E( E( Z_j \mid \bm A) )
    \\
    &=&
    \frac1{n-1} \sum_{j=2}^n E( A'^2_j + B'^2_j ),
    \\
    V(\hat\theta)
    &=&
    \frac1{n} E( V( X_1 \mid \bm A ) )
    \\
    &=&
    \frac1{n} E( A'^2_1+1/2 ) - E(\hat\theta)^2,
    \\
    V(\hat\eta)
    &=&
    \frac1{n} E( V( Y_1 \ \mid \bm A ) )
    \\
    &=&
    \frac1{n} E( B'^2_1+1/2 ) - E(\hat\eta)^2,
    \\
    V(\hat\nu)
    &=&
    \frac1{(n-1)^2} E\left( V\left( \left. \sum_{j=2}^{n} Z_j \ \right| \ \bm A \right) \right)
    \\
    &=&
    \frac1{(n-1)^2} \sum_{j=2}^{n} \left( E( (A'_j+B'_j)^4 + (A'_j+B'_j)^2 ) -  E(A'^2_j+B'^2_j)^2 \right)
    \\
    &&
    +
    \frac2{(n-1)^2} \sum_{2\le j<k\le n} \left( E( (A'_j+B'_j)^2 (A'_k+B'_k)^2 ) \right)
    \\
    &&
    -
    \frac2{(n-1)^2} \sum_{2\le j<k\le n} \left( E( (A'_j+B'_j)^2 ) E( (A'_k+B'_k)^2 ) \right)
    ,
  \end{eqnarray*}
  hold, so that these quantities depend only on
  bivariate, at most, fourth moments of $\bm A$,
  and we can obtain the following results.

\begin{theorem}\label{th:3:1}
  For any $\epsilon>0$, we have
  \begin{eqnarray}
    E(\hat\theta)
    &=&
    n^{-\epsilon/2} \theta,\quad
    E(\hat\eta)
    \ = \
    n^{-\epsilon/2} \eta
    ,
    \label{eq:3:E1}
    \\
    V(\hat\theta)
    &=&
    \frac{1 + \nu}{2 n}
     + \frac{\theta^2}{n} - \frac{ \theta^2}{n^{  \epsilon }}
    -\frac{ n^{-1} - n^{-\epsilon }}{2^{1+\epsilon} - 2^{2\epsilon}} \theta^2
    ,
    \label{eq:3:V1}
    \\
    V(\hat\eta)
    &=&
    \frac{1 + \nu}{2 n}
     + \frac{\eta^2}{n} - \frac{ \eta^2}{n^{  \epsilon }}
    -\frac{ n^{-1} - n^{-\epsilon }}{2^{1+\epsilon} - 2^{2\epsilon}} \eta^2
    ,
    \label{eq:3:V2}
    \\
    E(\hat\nu)
    &=&
    \nu
    +
    \frac{
      -1 + 2^{1 + \epsilon} - 2^{1 + \epsilon} n
      + 2^{2 \epsilon } n - 4^\epsilon + n^{ 1 - \epsilon }
    }
    {2^\epsilon (-2 + 2^\epsilon) (-1 + n)}
    (\theta^2+\eta^2)
    .
    \label{eq:3:E2}
  \end{eqnarray}
  If $\epsilon$ is positive and small enough, then we have
  \begin{equation}
    V(\hat\nu) \ = \ O(n^{-2\epsilon })
    \label{eq:3:V3}
  \end{equation}
  as $n$ goes to $\infty$.
\end{theorem}

{\bf PROOF.}\
  First, for $j=1,...,n\:(=2^m)$ and $t=0,1,...,m$,
    let $A_j^{(t)}$ and $B_j^{(t)}$ denote $A_j^{(t)}$ and $B_j^{(t)}$
    after the transformation $\prod_{(j,k)\in J_t} g_{\pi/4}^{(j,k)} \cdots \prod_{(j,k)\in J_1} g_{\pi/4}^{(j,k)}$
    which is on the way of $G_2$.
  Then, for any natural numbers $r_A$, $r_B$, $s_A$, $s_B$, $t$ and $j=2^{t-1}+1$,
    the recurrent equations
    \begin{eqnarray*}
      &&
      \hspace*{-1cm}
      E(A_1^{(t)r_A} B_1^{(t)r_B})
      \\
      &&
        = \
        E((A_1^{(t-1)}\cos\tau+A_j^{(t-1)}\sin\tau)^{r_A} (B_1^{(t-1)}\cos\tau+B_j^{(t-1)}\sin\tau)^{r_B})
        ,
      \\
      &&
      \hspace*{-1cm}
      E(A_1^{(t-1)r_A} B_1^{(t-1)r_B})
      \\
      &&
        = \
        E(A_j^{(t-1)r_A} B_j^{(t-1)r_B})
        .
      \\
      &&
      \hspace*{-1cm}
      E(A_1^{(t-1)r_A} B_1^{(t-1)r_B} A_j^{(t-1)s_A} B_j^{(t-1)s_B})
      \\
      &&
        = \
        E(A_1^{(t-1)r_A} B_1^{(t-1)r_B})
        \
        E(A_j^{(t-1)s_A} B_j^{(t-1)s_B})
    \end{eqnarray*}
    hold.
  Hence, we have (\ref{eq:3:E1}) to (\ref{eq:3:V2}).
  Next, for any $r_A$, $r_B$, $t$, $j=2^{t-1}+1$, and for any natural number $k$ satisfying
    \[
      j'-j \ \equiv \ 0 \pmod{2^t},
    \]
    the recurrent equations
    \begin{eqnarray*}
      &&
      \hspace*{-1cm}
      E(A_j^{(t)r_A} B_j^{(t)r_B})
      \\
      &&
        = \
        E((-A_1^{(t-1)}\sin\tau+A_j^{(t-1)}\cos\tau)^{r_A} (-B_1^{(t-1)}\sin\tau+B_j^{(t-1)}\cos\tau)^{r_B})
        ,
      \\
      &&
      \hspace*{-1cm}
      E(A_j^{(t)r_A} B_j^{(t)r_B}) \ = \ E(A_{j'}^{(t)r_A} B_{j'}^{(t)r_B})
    \end{eqnarray*}
    hold.
  Hence, we have (\ref{eq:3:E2}).
  Finally, for any $r_A$, $r_B$, $s_A$, $s_B$, $t$, $\Delta=1,...,m-t$,
    $j=2^{t-1}+1$ and $k=2^{t+\Delta-1}+1$,
    the recurrent equations
    \begin{eqnarray*}
      &&
      \hspace*{-1cm}
      E( A_1^{(t)r_A} B_1^{(t)r_B} A_j^{(t)s_A} B_j^{(t)s_B})
      \\
      &&
        = \
        E(
        (A_1^{(t-1)}\cos\tau + A_j^{(t-1)}\sin\tau)^{r_A}
        (B_1^{(t-1)}\cos\tau + B_j^{(t-1)}\sin\tau)^{r_B}
      \\
      &&
        \qquad\times
        (-A_1^{(t-1)}\sin\tau + A_j^{(t-1)}\cos\tau)^{s_A}
        (-B_1^{(t-1)}\sin\tau + B_j^{(t-1)}\cos\tau)^{s_B}
        )
        ,
      \\
      &&
      \hspace*{-1cm}
      E( A_1^{(t+\Delta)r_A} B_1^{(t+\Delta)r_B} A_j^{(t)s_A} B_j^{(t)s_B} )
      \\
      &&
        = \
        E(
        (A_1^{(t+\Delta-1)}\cos\tau + A_k^{(t+\Delta-1)}\sin\tau)^{r_A}
      \\
      &&
        \qquad\times
        (B_1^{(t+\Delta-1)}\cos\tau + B_k^{(t+\Delta-1)}\sin\tau)^{r_B}
      \\
      &&
        \qquad\times
        (-A_1^{(t-1)}\sin\tau + A_j^{(t-1)}\cos\tau)^{s_A}
        (-B_1^{(t-1)}\sin\tau + B_j^{(t-1)}\cos\tau)^{s_B}
        )
        ,
      \\
      &&
      \hspace*{-1cm}
      E( A_k^{(t+\Delta)r_A} B_k^{(t+\Delta)r_B} A_j^{(t)s_A} B_j^{(t)s_B} )
      \\
      &&
        = \
        E(
        (-A_1^{(t+\Delta-1)}\sin\tau + A_k^{(t+\Delta-1)}\cos\tau)^{r_A}
      \\
      &&
        \qquad\times
        (-B_1^{(t+\Delta-1)}\sin\tau + B_k^{(t+\Delta-1)}\cos\tau)^{r_B}
      \\
      &&
        \qquad\times
        (-A_1^{(t-1)}\sin\tau + A_j^{(t-1)}\cos\tau)^{s_A}
        (-B_1^{(t-1)}\sin\tau + B_j^{(t-1)}\cos\tau)^{s_B}
        )
    \end{eqnarray*}
    hold.
  Hence, we have (\ref{eq:3:V3}).
\hfill$\Box$

By this theorem, we can see that Hayashi's estimators
  $\hat\theta$, $\hat\eta$ and $\hat\nu$ are biased
  and asymptotically inconsistent.

\begin{figure}[p]
    \hspace*{-2cm}
    \rotatebox{90}
    {
    \begin{minipage}{\textheight}
    \begin{center}
      \includegraphics{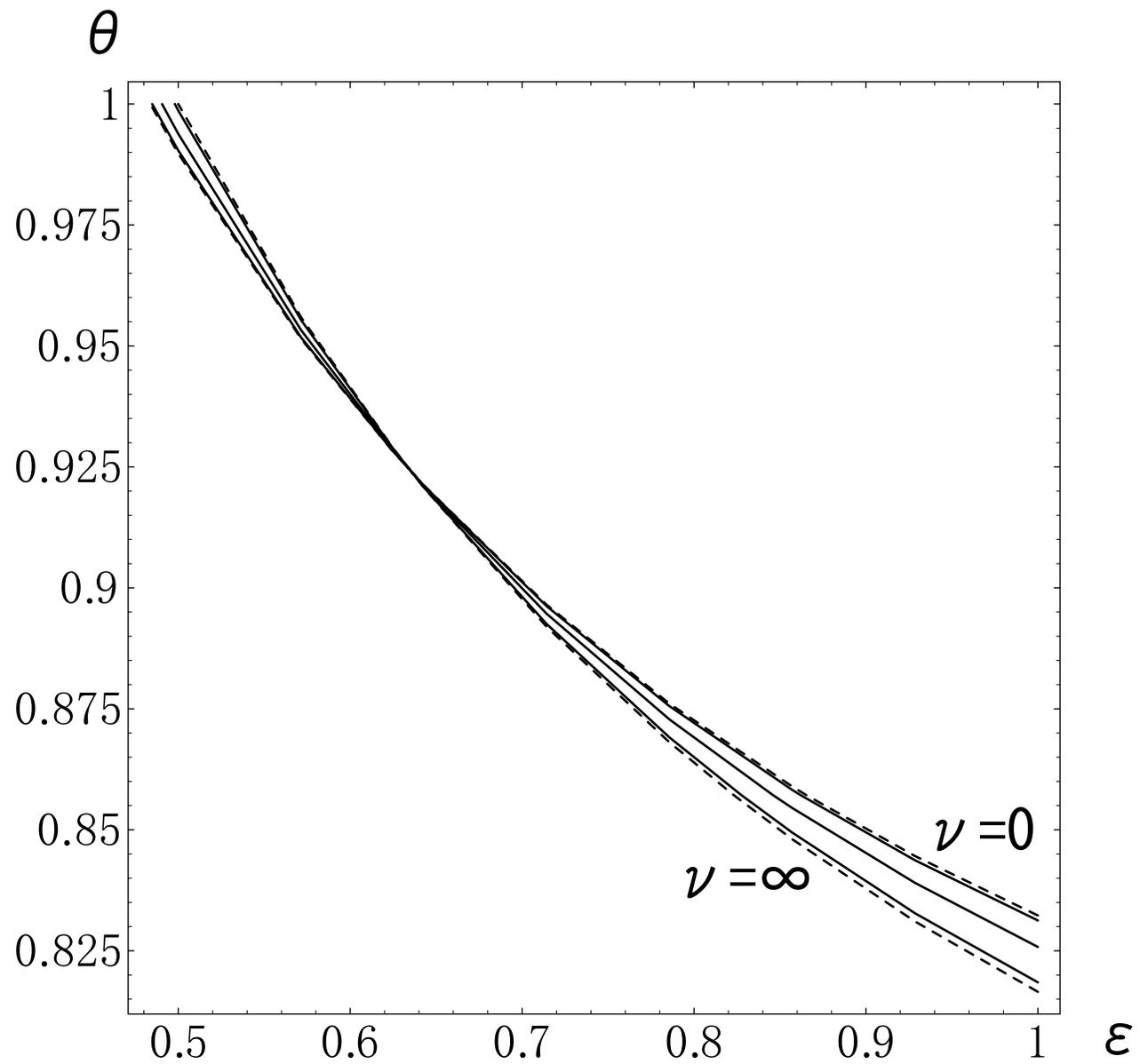}
    \end{center}
    \caption
    {
      Pairs of $(\epsilon,\theta)$ satisfying
      $\hat M=\bar M$ for $\nu=0.1,1,10,100$ (solid lines)
     and for $\nu=0,\infty$ (dashed lines) in case of $\eta=0$.
      \label{fig:3:gaugra1}
    }
    \end{minipage}
    }
  \end{figure}

Let us compare $\bar\nu$ and $\hat\nu$.
Since $\hat\nu$ is biased, it is better to compare
  them by mean square errors, that is,
  \begin{eqnarray*}
    \bar M & := & E((\bar\nu-\nu)^2) \ = \ V(\hat\nu)
    ,
    \\
    \hat M & := & E((\hat\nu-\nu)^2) \ = \ V(\hat\nu)+E(\hat\nu-\nu)^2
    .
  \end{eqnarray*}
If $n=2$, then we have
  \begin{eqnarray*}
    \bar M
    &=&
    (\nu+1)^2
    ,
    \\
    \hat M
    &=&
    \nu^2 + \nu
    + (2\nu+1) ( 1 - 2^{ - 2 \epsilon} ) (\theta^2 + \eta^2 )
    +
    \frac{ 3 + 2^{-8 \epsilon} - 4^{1 - \epsilon}   }{2}
    (\theta^4+\eta^4)
    .
  \end{eqnarray*}
We can see that, if $\epsilon$ or $\theta^2+\eta^2$ is small enough,
  then $\hat M<\bar M$.
In case of $\eta=0$, $(\epsilon,\theta)$ satisfying
  $\hat M=\bar M$ are numerically calculated for $\nu=0.1,1,10,100$ (solid lines)
  and for $\nu=0,\infty$ (dashed lines) and plotted in Figure \ref{fig:3:gaugra1}.
If the parameters are in the lower-left side of the lines,
  then $\hat M<\bar M$, so that Hayashi's estimator $\nu$ for
  $\nu$ has advantage in the sense of mean square error.

\bigskip
\hfil\fbox{{\Huge\bf Figure \ref{fig:3:gaugra1}}}

\

\section{Correcting Hayashi's method by stopping the transformation}
We have seen that Hayashi's estimators $\hat\theta$, $\hat\eta$ and $\hat\nu$
  are not so good asymptotically if the transformation is noisy.
Hence, in the case of $n=2^m$, we consider to stop
  the transformation $G_2$ of Remark \ref{rm:3:g2} at $m_0$-th step.
Let $W'_j:=(A'_j,B'_j,X_j,Y_j,Z_j)$ $(j=1,...,n)$ be random variables
  obtained by the transformation
  \[
    \prod_{(j,k)\in J_{m_0}} g_{\tau}^{(j,k)} \cdots \prod_{(j,k)\in J_t} g_{\tau}^{(j,k)} \cdots \prod_{(j,k)\in J_1} g_{\tau}^{(j,k)}
  \]
  of $W_j:=(A_j,B_j,X_j,Y_j,Z_j)$ $(j=1,...,n)$.
Let
  \[
    s \ := \ \{ 2^{m_0} j + 1 \mid j=0,...,2^{m-m0}-1 \}
  \]
  be a set of indices of which
  we observe $(X_j,Y_j)$, hence we observe $Z_k$ if $k\not\in S$.
Let
  \begin{eqnarray*}
    \tilde\theta_{m_0}
    &:=&
    \frac{1}{2^{m-m_0/2-\epsilon m_0/2}}\sum_{j\in S} X_j
    ,
    \quad
    \tilde\eta_{m_0}
    \ := \
    \frac{1}{2^{m-m_0/2-\epsilon/2}}\sum_{j\in S} Y_j
    ,
    \\
    \tilde\nu_{m_0}
    &:=&
    \frac1{2^m-1} \sum_{k\not\in S} Z_k
    +
    \frac{2^{m-m_0} - 1 }{(2^m-1) 2^{ m-m_0 }}
    \sum_{j\in S} (X^2_j+Y^2_j)
    \\
    &&
    -\frac\alpha\beta \sum_{j,k\in S,j\ne k} (X_jX_k+Y_jY_k)
    -\frac{2^m - 2^{m_0}}{2^{m_0} (  2^m -1)}
  \end{eqnarray*}
  where
  \begin{eqnarray*}
    \alpha
    &:=&
    2^{-\epsilon - m -  m_0 }
     \Big( 2^{( 2 + \epsilon )   m_0 } - 
     2^{1 + \epsilon + 2  m_0  + 
     \epsilon  m_0 } + 
     2^{1 + \epsilon + m + 2  m_0  + 
     \epsilon  m_0 }
  \\
  &&
    + 2^{2  m_0  +
    \epsilon ( 2 +  m_0  ) } - 
    2^{m + 2  m_0  + 
    \epsilon ( 2 +  m_0  ) } - 
    8^{ m_0 } \Big)
    ,
    \\
    \beta
    &:=&
    ( -2 + 2^\epsilon )  ( -1 + 2^m ) ( -2^m + 2^{ m_0 } )
    .
  \end{eqnarray*}
Then, by the same argument of the proof of Theorem \ref{th:3:1},
  we have
  \begin{eqnarray*}
    E(\tilde\theta_{m_0}) & = & \theta,\quad
    E(\tilde\eta_{m_0})   \ = \ \eta,
    \\
    V(\tilde\theta_{m_0})
    &=&
    \frac{1+\nu}{2^{ 1 + m -e m_0 } }
    -\frac{ ( 2^e -1 )^2 (2^{m_0} - 2^{e m_0})}
      {  2^{e + m} (2^e -2 ) } \theta^2
      ,
    \\
    V(\tilde\eta_{m_0})
    &=&
    \frac{1+\nu}{2^{ 1 + m -e m_0 } }
    -\frac{ ( 2^e -1 )^2 (2^{m_0} - 2^{e m_0})}
      {  2^{e + m} (2^e -2 ) } \eta^2
      ,
  \end{eqnarray*}
  and, if $m_0$ is fixed, then
  \[
    V(\tilde\nu_{m_0})
    \ = \
    O(n^{-1})
  \]
  as $n$ goes to $\infty$.

\section{Numerical comparison and conclusion}
For $n=2^6,2^8,2^{10},2^{12}$ and for $\epsilon=0.0001$,
  $\tilde\theta_{m_0}$, $\hat\eta_{m_0}$ and $\hat\nu_{m_0}$ are constructed,
  and, for $\theta=0,2,4,6,8,10$, $\eta=0$ and $\nu=0,2,4,6,8,10$,
  the sum of mean square errors
  \[
    M_{m_0} \ := \ E( (\tilde\theta_{m_0}-\theta)^2
                     +(\tilde\eta_{m_0}-  \eta)^2
                     +(\tilde\nu_{m_0}-\nu)
                    )
  \]
  are numerically calculated and their relative errors
  \[
    1-M_{m_0}/M_0
  \]
  are shown in Table 1.
The Hayashi's estimators or the corrected estimators are better than the
	naive estimators if the relative error is positive.
The corrected estimators $\tilde\theta_{m/2}$, $\tilde\eta_{m/2}$ and
	$\tilde\nu_{m/2}$ are better than the Hayashi's estimators if the
	relative error of the corrected ones is larger.

\begin{figure}
\begin{center}
	\scalebox{1.5}{\includegraphics{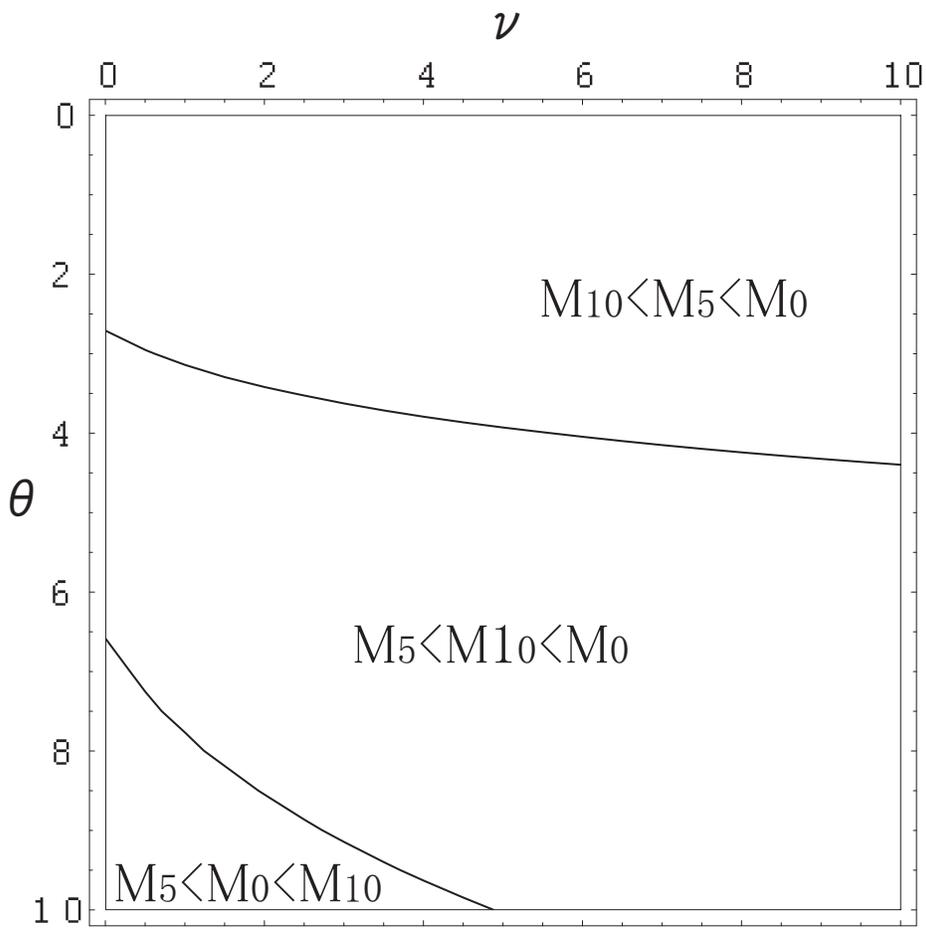}}
	\caption{\label{fig:hayashivstsuda}
	The goodness of estimators for $n=2^{10}$.}
\end{center}
\end{figure}

When the location parameter $\theta$ is close to the origin or
	the scale parameter $\nu$ is large,
	the effect of the randomness of the transformation is small
	and Hayashi's estimators has the good performance.
However,  when the location is far from the origin or
	the scale is small, the randomness of the transformations
	significantly influences Hayashi's estimators and its performance
	is inferior to the naive estimator.
Our corrected estimators also loses the accuracy when the location is far from
	the origin or the scale is small, but such bad effects are relatively
	smaller than that of Hayashi's estimators.
Indeed, in the range of our simulation, the corrected estimators are always
	better than the naive estimators
	(see Figure \ref{fig:hayashivstsuda} for $n=2^{10}$).

\bigskip\hfil\fbox{{\Huge Table 1}}\bigskip

\bigskip\hfil\fbox{{\Huge Figure \ref{fig:hayashivstsuda}}}\bigskip

\bigskip


\begin{flushleft}
{\large\bf Acknowledgments}
\end{flushleft}

The authors wish to thank
  Professor M. Akahira of the University of Tsukuba
  and
  Professor H. Imai of the University of Tokyo
  for their support and encouragement.
The authors wish to thank
  Dr. A. Tomita and Dr. F. Yura
  of IMAI Quantum Computation and Information Project
  for useful comments from physical points of view.

\renewcommand{\refname}{References}

\pagestyle{empty}
\include{tabledec}

\end{document}